\documentclass[sigconf]{acmart}
\usepackage{flushend}
\hypersetup{draft}

\AtBeginDocument{%
  }

\copyrightyear{2025}
\acmYear{2025}
\setcopyright{cc}
\setcctype{by-nc}
\acmDOI{10.1145/3696673.3723083}

\acmConference[ACMSE 2025]{2025 ACM Southeast Conference}{April 24--26,
2025}{Cape Girardeau, MO, USA}
\acmBooktitle{2025 ACM Southeast Conference (ACMSE 2025), April 24--26, 2025,
Cape Girardeau, MO, USA}
\acmISBN{979-8-4007-1277-7/2025/04}

\usepackage{algorithm}
\usepackage{algpseudocode}
\usepackage{algorithmicx}

\begin{document}
\pagestyle{empty}

\title{Improving Merge Sort and Quick Sort Performance by Utilizing Alphadev’s Sorting Networks as Base Cases}

\author{Anas Gamal Aly}
\affiliation{%
  \institution{Stetson University}
  \city{DeLand}
  \state{Florida}
  \country{USA}
}
\email{agamal@stetson.edu}

\author{Anders E. Jensen}
\affiliation{%
  \institution{Stetson University}
  \city{DeLand}
  \state{Florida}
  \country{USA}
}
\email{aejensen@stetson.edu}

\author{Hala ElAarag}
\affiliation{%
  \institution{Stetson University}
  \city{DeLand}
  \state{Florida}
  \country{USA}
}
\email{helaarag@stetson.edu}

\renewcommand{\shortauthors}{Aly et al.}

\begin{abstract}
Recent work by Google DeepMind introduced assembly-optimized sorting networks that achieve faster performance for small fixed-size arrays (3-8). In this research, we investigate the integration of these networks as base cases in classical divide-and-conquer sorting algorithms, specifically Merge Sort and Quick Sort, to leverage these efficient sorting networks for small subarrays generated during the recursive process. We conducted benchmarks with 11 different optimization configurations and compared them to classical Merge Sort and Quick Sort. We tested the configurations with random, sorted and nearly sorted arrays.

Our optimized Merge Sort, using a configuration of three sorting networks (sizes 6, 7, and 8), achieves at least 1.5x speedup for random and nearly sorted arrays, and at least 2x speedup for sorted arrays, in comparison to classical Merge Sort. This optimized Merge Sort surpasses both classical Quick Sort and similarly optimized Quick Sort variants when sorting random arrays of size 10,000 and larger.

When comparing our optimized Quick Sort to classical Quick Sort, we observe a 1.5x speedup using the 3-to-5 configuration on sorted arrays of size 10,000. The 6-to-8 configuration maintains a consistent 1.5x improvement across sorted arrays from 25,000 to 1 million elements. Our findings demonstrate the potential of integrating AI-optimized sorting networks to enhance the performance of classical sorting algorithms.

\end{abstract}



\maketitle

\section{Introduction}
Sorting algorithms form a critical foundation of modern computing infrastructure, powering numerous applications from online shopping and social media feeds to scientific research and financial analysis. While classical sorting algorithms like merge sort achieve \begin{math}O(n\log n)\end{math} complexity, there remains significant potential for practical performance improvements. Recent research by Google DeepMind introduced AlphaDev~\cite{AlphaDev2023}, which discovered novel sorting networks optimized at the assembly level for small fixed-size arrays. These networks cover array sizes from 3 to 8 elements, presenting an opportunity to enhance classical sorting algorithms.

Our research investigates the integration of these fixed sorting networks as optimized base cases within classical sorting algorithms. We focus on Merge Sort and Quick Sort, implementing and evaluating hybrid approaches that leverage AlphaDev's sorting networks at different array size thresholds. This work bridges the gap between theoretical algorithmic improvements and practical performance optimization by demonstrating how assembly-level optimizations can enhance traditional divide-and-conquer sorting strategies. Source code is available online on GitHub\footnote{https://github.com/anasgamal/alphadev-merge-quick-sort}.

The contributions of this paper include: (1) a systematic evaluation of different sorting network configurations as base cases for classical sorting algorithms, particularly Merge Sort and Quick Sort, (2) empirical evidence showing that our optimized Merge Sort implementation achieves a 2x speedup for random arrays of size 10,000 and maintains a 1.5x improvement for arrays up to 1 million elements using the 6-to-8 configuration, while our enhanced Quick Sort demonstrates a 1.5x speedup on sorted arrays using the 3-to-5 configuration, and (3) a practical demonstration of how AI-optimized assembly code, specifically AlphaDev's 
~\cite{AlphaDev2023} sorting networks, can enhance traditional algorithms while maintaining their general applicability across different input sizes and distributions.

\section{Related Works}
Our research builds upon several key contributions in sorting algorithm optimization and artificial intelligence. In the domain of sorting algorithms, Pandey and Gupta's work~\cite{pandey2024lazy} on Lazy Merge Sort established that \begin{math}O(n\log n)\end{math} complexity serves as an upper bound for potential improvements to the classical merge sort algorithm. While Marszalek's~\cite{sym9090176} approach in 2017 focused on leveraging modern architecture and parallelization techniques, our work explores a novel direction by incorporating AI-generated sorting networks.

Recent developments in Artificial Intelligence (AI), particularly in Large Language Models (LLMs), have sparked increased interest in utilizing computational intelligence for various tasks, including code generation and optimization~\cite{zhao2023survey}. Reinforcement Learning (RL), a technique where agents learn optimal behaviors through reward systems~\cite{sun2023reinforcement}, has been particularly influential in improving LLMs' ability to follow user instructions~\cite{sun2023reinforcement}. The emergence of "LLM-based autonomous agents" has demonstrated advantages over traditional RL agents, even without domain-specific knowledge~\cite{wang2024survey}.

The foundation of our implementation relies on the groundbreaking work by Mankowitz et al.~\cite{AlphaDev2023}, who demonstrated how deep reinforcement learning could optimize sorting algorithms at the CPU instruction level. Their research introduced sorting networks optimized for actual measured latency at the assembly level, which we utilize in our work. Our approach extends these findings by systematically integrating these networks into classical sorting algorithms and examining their effectiveness across varying input sizes and patterns.

Our work utilizes AlphaDev~\cite{AlphaDev2023}, an RL-based system (distinct from LLM-based approaches), to improve classical sorting algorithms through assembly-optimized sorting networks. This implementation demonstrates how targeted AI applications can enhance well-established algorithmic methods through low-level optimizations.

\section{Methodology}

Our implementation utilized C++ to interface with AlphaDev's assembly-level sorting networks. Testing was conducted on a Windows 11 PC using Ubuntu WSL. The testing environment specifications are outlined in Table~\ref{tab:setup}. Our implementation process began with establishing baseline performance using classical merge sort and quick sort algorithms. We then integrated sorting networks from the AlphaDev GitHub Repository~\cite{alphadev_repo} to create optimized versions of both algorithms.

\begin{table}
\caption{System Configuration for Performance Testing}
\label{tab:setup}
\begin{tabular}{ll}
\toprule
\textbf{Component} & \textbf{Specification} \\
\midrule
CPU Model & AMD Ryzen 5 1600AF \\
CPU Clock Speed & 3.2 GHz \\
CPU Cores/Threads & 6/12 \\
RAM & 16 GB DDR4 \\
Architecture & AMD64 \\
Operating System & Windows 11 (WSL) \\
Compiler & Ubuntu clang 18.1.3 \\
Build System & Bazel 7.4.1 \\
Testing Framework & Google Test 1.12.1 \\
Benchmarking Tool & Google Benchmark 1.5.0 \\
\bottomrule
\end{tabular}
\end{table}

The key innovation in our approach is the integration of sorting networks as optimized base cases within the classical merge sort algorithm. Rather than continuing recursive division until reaching single elements, we check if a subarray's size matches any of our available sorting networks. If there is a match, we apply the corresponding network directly. Algorithm ~\ref{alg:optimizedmergesort} demonstrates this hybrid approach.

\begin{algorithm}
\caption{Optimized Merge Sort with Sorting Networks}\label{alg:optimizedmergesort}
\begin{algorithmic}[1]
\Function{OptimizedMergeSort}{array, left, right}
    \State size = right - left + 1
    \If{size matches sorting network size}
        \State ApplySortingNetwork(array[left...right])
        \State \Return
    \EndIf
    \If{left < right}
        \State mid = left + (right - left) / 2
        \State OptimizedMergeSort(array, left, mid)
        \State OptimizedMergeSort(array, mid + 1, right)
        \State Merge(array, left, mid, right)
    \EndIf
\EndFunction
\end{algorithmic}
\end{algorithm}

For example, in our 6-to-8 configuration, when the algorithm encounters a subarray of size 6, 7, or 8, it directly applies the corresponding AlphaDev~\cite{AlphaDev2023} sorting network instead of further recursion. This optimization reduces the number of comparisons and memory operations for these critical base cases while maintaining the algorithm's overall divide-and-conquer structure for larger arrays.

Similarly, our Quick Sort optimization integrates sorting networks for small subarrays, as shown in Algorithm~\ref{alg:optimizedquicksort}. However, Quick Sort's approach differs in two key aspects: it uses the median-of-three method for pivot selection and Hoare's partitioning scheme~\cite{10.1093/comjnl/5.1.10}, which together with the sorting networks creates a more sophisticated hybrid algorithm. When a partition falls within the size range of available sorting networks (for example, 6-8 elements in the 6To8 configuration), the algorithm bypasses further partitioning and applies the appropriate sorting network directly.

\begin{algorithm}[H]
\caption{Optimized Quick Sort with Sorting Networks}\label{alg:optimizedquicksort}
\begin{algorithmic}[1]
\Function{OptimizedQuickSort}{array, low, high}
    \State size = high - low + 1
    \If{size matches sorting network size}
        \State ApplySortingNetwork(array[low...high])
        \State \Return
    \EndIf
    \If{low < high}
        \State pivotIndex = MedianOfThree(array, low, high)  \Comment{Pivot selection}
        \State partitionIndex = HoarePartition(array, low, high, pivotIndex)
        \State OptimizedQuickSort(array, low, partitionIndex)
        \State OptimizedQuickSort(array, partitionIndex + 1, high)
    \EndIf
\EndFunction
\end{algorithmic}
\end{algorithm}

We conducted benchmarking using Google Benchmark~\cite{googlebenchmark}, which automatically determines the number of iterations needed to achieve statistical significance. For example, when testing MergeSortClassic with random arrays of size 10,000, the framework ran 646 iterations to ensure reliable timing measurements. For larger arrays of size 25,000 and 1 million, it ran 261 and 6 iterations respectively, adjusting the iteration count to maintain measurement quality while accounting for longer execution times. Results were collected and exported to CSV format for analysis and visualization.

\begin{table}
\caption{Testing Parameters and Data Generation Specifications}
\label{tab:testing-params}
\begin{tabular}{ccl}
\toprule
\textbf{Parameter} & \textbf{Details} \\
\midrule
Input Range & 10000 to 1000000 \\
Distributions & Random, sorted, nearly sorted \\
Framework & Google Benchmark 1.5.0 \\
Measurements & CPU and wall time \\
Iterations & Auto-scaled by benchmark \\
\bottomrule
\end{tabular}
\end{table}

\section{Results}

For each algorithm we tested, we had 12 implementations including a classical implementation of such algorithms without our optimizations. The other 11 implementations described in Table  ~\ref{tab:configs} are optimizations that have different configurations utilizing either fixed or dynamic sorting networks as base cases. VarSort3/4/5 refers to dynamic sorting networks that can handle variable-sized inputs up to sizes 3, 4, and 5 respectively. Speedup factors (Eq. \ref{eq:speedup}) represent the ratio of classical to optimized execution times. We selected the best performer configurations for Merge Sort and Quick Sort for the figures in this section. We chose VarSort4 to demonstrate as a representative of dynamic sorting networks performance, since most of them did not provide significant performance improvements.
\begin{equation}
\text{Speedup factor} = T_{\text{classical}} / T_{\text{optimized}}
\label{eq:speedup}
\end{equation}
\begin{table}
  \caption{Sorting Network Configurations}
  \label{tab:configs}
  \begin{tabular}{ccl}
    \toprule
    Configuration & Networks Used\\
    \midrule
    PowerOf2 & 4, 8\\
    Even & 4, 6, 8\\
    Odd & 3, 5, 7\\
    3 & 3 \\
    3To4 & 3, 4\\
    3To5 & 3, 4, 5\\
    3To8 & 3, 4, 5, 6, 7, 8\\
    6To8 & 6, 7, 8 \\
    VarSort3 & VarSort3\\
    VarSort4 & VarSort4\\
    VarSort5 & VarSort5\\

    \bottomrule
  \end{tabular}
\end{table}

\subsection{Merge Sort}

This section analyzes the performance improvements achieved by integrating AlphaDev's sorting networks into the Merge Sort algorithm. We evaluate the speedup attained by different configurations (3-to-5, 3-to-8, 6-to-8, PowerOf2, and VarSort4) across various array sizes and input characteristics: random arrays, sorted arrays, and nearly sorted arrays

\subsubsection{Random Arrays}
\label{sub:mergesort-random}

\begin{figure}[h]
  \centering
  \includegraphics[width=\linewidth]{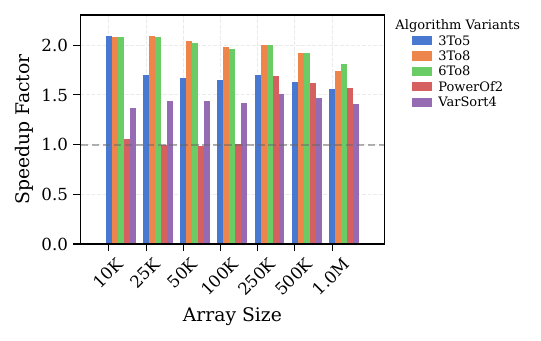}
  \caption{Merge Sort Speed Up Analysis with Random Arrays}
  \Description{Plot demonstrating speed up in merge sort sorting random arrays with our optimizations}
  \label{fig:mergesort-random-speedup}
\end{figure}

Figure~\ref{fig:mergesort-random-speedup} illustrates the performance gains achieved by integrating AlphaDev's~\cite{AlphaDev2023} sorting networks into Merge Sort for random arrays. Notably, the 6-to-8 configuration, employing only sorting networks of sizes 6, 7, and 8, consistently delivers substantial speedups across all tested array sizes (10,000 to 1 million).

While the 3-to-5 configuration shows a near 2x speedup for arrays of size 10,000, its advantage diminishes beyond 25,000. The 3-to-8 and 6-to-8 configurations maintain near 2x speedups up to arrays of size 50,000, and although they dip slightly at arrays of size 100,000, they rebound to 2x at arrays of size 250,000.  For larger arrays (approaching 1 million), the speedup converges towards 1.5x, with the 6-to-8 configuration consistently outperforming 3-to-8, highlighting its scalability.

The 6-to-8 configuration offers an optimal balance of simplicity and effectiveness. Its use of only three sorting networks provides a remarkably stable speedup across the entire array size range, making it the most practical choice for optimizing Merge Sort on random arrays. The minor performance fluctuation at arrays of size 100,000 is insignificant compared to its overall consistency and simplicity.
\subsubsection{Sorted Arrays}

\begin{figure}[h]
  \centering
  \includegraphics[width=\linewidth]{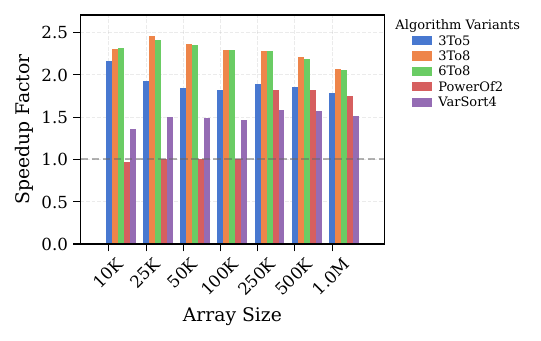}
  \caption{Merge Sort Speed Up Analysis with Sorted Arrays}
  \Description{Plot demonstrating speed up in merge sort sorting sorted arrays with our optimizations}
  \label{fig:mergesort-sorted-speedup}
\end{figure}

For already sorted arrays, Figure~\ref{fig:mergesort-sorted-speedup} reveals that the 3-to-8 configuration provides the most significant and consistent performance improvement. It maintains a speedup slightly above 2x across all tested array sizes, from 10,000 to 1 million. The 6-to-8 configuration exhibits similar performance, closely tracking the 3-to-8 results, albeit with a marginally lower speedup.

The 3-to-5 configuration, while initially comparable to 3-to-8 and 6-to-8 for 10,000 arrays, shows a noticeable decline in performance as array size increases. Its speedup drops below 2x for arrays of 25,000 and larger, indicating that it's less effective for larger, sorted datasets in the context of Merge Sort.

It is noteworthy that the performance gains observed on sorted arrays are, overall, more pronounced than those seen on random arrays (as discussed in Section~\ref{sub:mergesort-random}). 6-to-8 remains a good configuration considering its simplicity. 

\subsubsection{Nearly Sorted Arrays}

\begin{figure}[h]
  \centering
  \includegraphics[width=\linewidth]{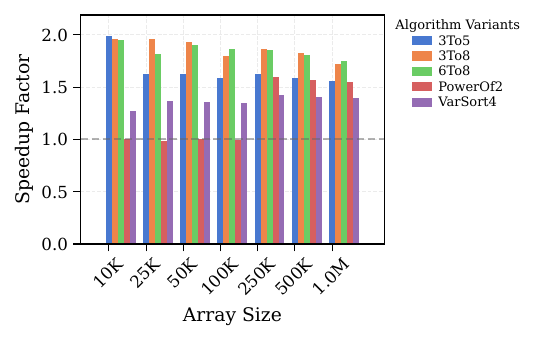}
  \caption{Merge Sort Speed Up Analysis with Nearly Sorted Arrays}
  \Description{Plot demonstrating speed up in merge sort sorting nearly sorted arrays with our optimizations}
  \label{fig:mergesort-nearly-sorted-speedup}
\end{figure}

Figure \ref{fig:mergesort-nearly-sorted-speedup} illustrates the speedup achieved by various merge sort optimizations when applied to nearly sorted arrays of different sizes. As shown, the 3-to-5 variant achieved the highest speedup, approaching a 2x improvement, for arrays with 10,000 elements. However, its performance degraded relative to 3-to-8 and 6-to-8 as the array size increased beyond 25,000.

The 3-to-8 variant demonstrated the most consistent performance across all array sizes, maintaining a speedup factor of over 1.5. The 6-to-8 variant exhibited comparable performance to 3-to-8 for most array sizes, even slightly outperforming it for arrays with 1 million elements. While 3-to-8 emerges as the most consistently performant algorithm, 6-to-8 presents a viable alternative for nearly sorted arrays, particularly due to its simpler implementation. The PowerOf2 and VarSort4 variants show speedups for some array size and input type combinations but are typically outperformed by the 3-to-8 and 6-to-8 variants.

\subsection{Quick Sort}

Quick Sort was also implemented with the same configurations illustrated in Table ~\ref{tab:configs} that we used in our Merge Sort experiments.

\subsubsection{Random Arrays}

\begin{figure}[h]
  \centering
  \includegraphics[width=\linewidth]{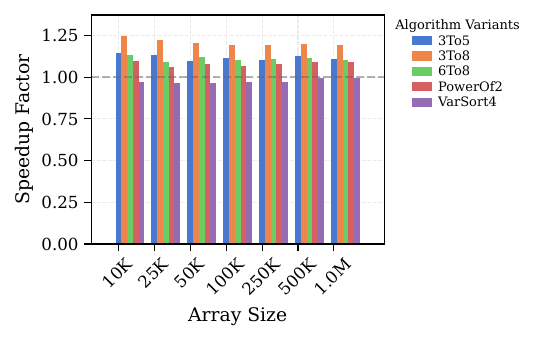}
  \caption{Quick Sort Speed Up Analysis with Random Arrays}
  \Description{Plot demonstrating speed up in quick sort sorting random arrays with our optimizations}
  \label{fig:quicksort-random-speedup}
\end{figure}

Figure ~\ref{fig:quicksort-random-speedup} shows that for random arrays, the 3-to-8 configuration achieves the most significant speedup, reaching approximately 1.25x for arrays of size 10,000.  However, this improvement diminishes as the array size increases.  The 3-to-5 and 6-to-8 configurations show negligible speedup across all array sizes. This suggests that for Quick Sort with random input, the integration of AlphaDev's sorting networks offers limited performance benefits.

\subsubsection{Sorted Arrays}

\begin{figure}[h]
  \centering
  \includegraphics[width=\linewidth]{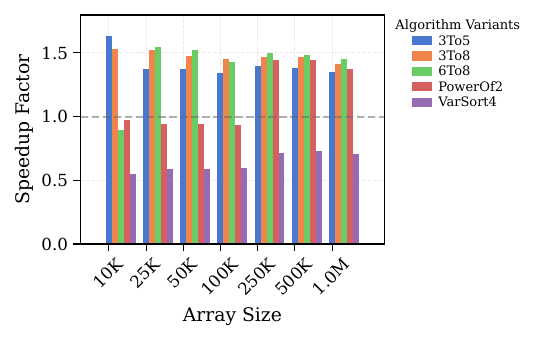}
  \caption{Quick Sort Speed Up Analysis with Sorted Arrays}
  \Description{Plot demonstrating speed up in quick sort sorting sorted arrays with our optimizations}
  \label{fig:quicksort-sorted-speedup}
\end{figure}

Figure ~\ref{fig:quicksort-sorted-speedup} illustrates the speedup achieved by Quick Sort optimizations on sorted arrays. Notably, the 3-to-5 configuration yields the highest speedup for arrays of size 10,000, reaching 1.5x. However, as the array size increases, the 6-to-8 configuration becomes more effective. For arrays of size 25,000 and larger, the 6-to-8 configuration consistently provides a speedup of around 1.5x, maintaining this improvement even for arrays with 1 million elements. This suggests that while 3-to-5 is optimal for smaller sorted arrays, 6-to-8 provides more consistent benefits as the input size grows larger than 25,000.

\subsubsection{Nearly Sorted Arrays}

\begin{figure}[h]
  \centering
  \includegraphics[width=\linewidth]{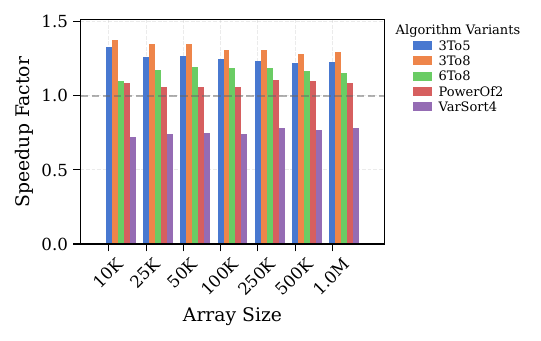}
  \caption{Quick Sort Speed Up Analysis with Nearly Sorted Arrays}
  \Description{Plot demonstrating speed up in quick sort sorting nearly sorted arrays with our optimizations}
  \label{fig:quicksort-nearly-sorted-speedup}
\end{figure}

Figure ~\ref{fig:quicksort-nearly-sorted-speedup}  shows the speedup of Quick Sort optimizations on nearly sorted arrays. Similar to the trend observed with sorted arrays, the 3-to-5 configuration provides the highest speedup for smaller array sizes (around 1.4x for 10,000 elements). However, the 3-to-8 configuration takes the lead in overall performance across all array sizes, consistently achieving a speedup of approximately 1.3x for arrays with 10,000 elements and above.  While the 6-to-8 configuration offers minimal speedup for arrays with 10,000 elements, it becomes more effective with larger array sizes (25,000 and above).  Despite this improvement, it consistently falls behind the 3-to-8 configuration in terms of speedup.

\subsection{Merge Sort vs Quick Sort}

\begin{figure*}[h]
  \centering
  \includegraphics[width=\linewidth]{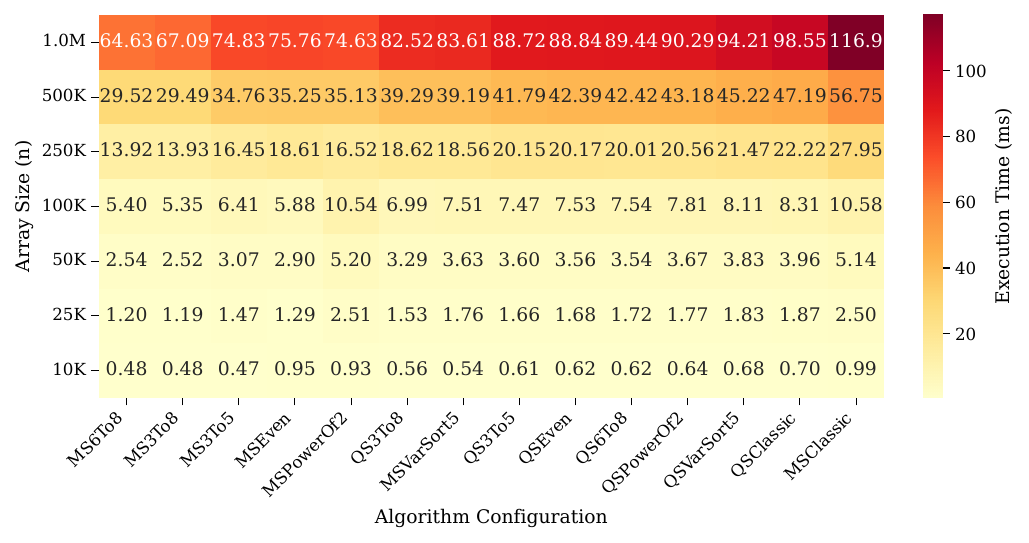}
  \caption{Merge Sort (MS) vs Quick Sort (QS) Comparison Across Different Configurations with Random Arrays}
  \Description{Heatmap comparison of Merge Sort and Quick Sort variants.}
  \label{fig:qsvsmsheatmap-random}
\end{figure*}

While previous research by Taiwo et al.~\cite{taiwo2020comparative} demonstrated that classical Quick Sort outperforms classical Merge Sort for larger arrays, our findings reveal a nuanced picture when incorporating AlphaDev's~\cite{AlphaDev2023} sorting networks. As shown in Figure ~\ref{fig:qsvsmsheatmap-random}, we confirm the previous results for classical implementations with array sizes of 10,000 and larger.

However, our experiments demonstrate that certain optimized Merge Sort variants surpass both classical Quick Sort and similarly optimized Quick Sort variants, particularly when sorting random arrays.  Specifically, the Merge Sort 6-to-8 configuration, utilizing only three sorting networks (sizes 6, 7, and 8), achieves a remarkable 1.5x speedup over classical Quick Sort and a 1.8x speedup over classical Merge Sort. This makes our optimized Merge Sort with the 6-to-8 configuration a compelling choice for applications frequently dealing with large random arrays, outperforming both classical and optimized Quick Sort implementations.

\begin{figure*}[h]
  \centering
  \includegraphics[width=\linewidth]{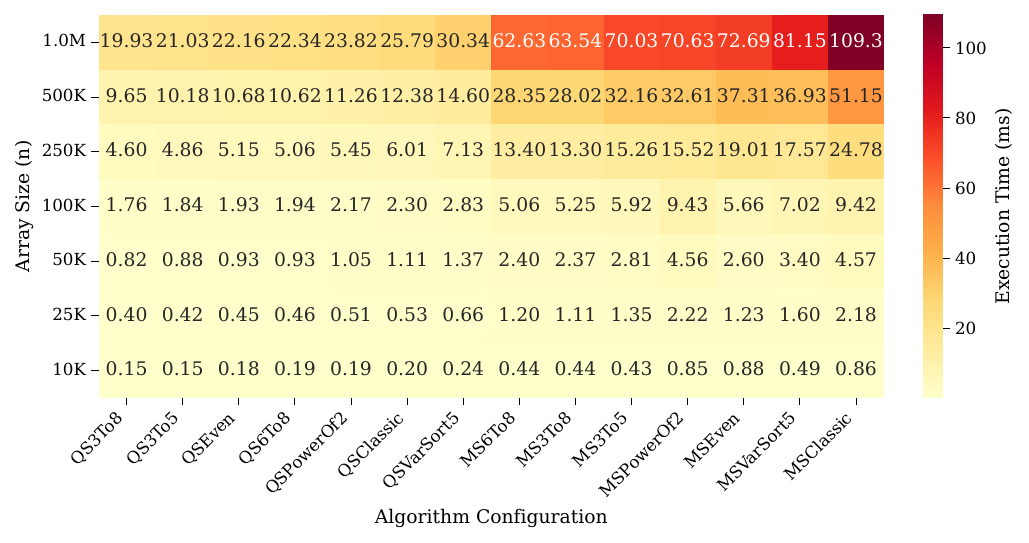}
  \caption{Merge Sort (MS) vs Quick Sort (QS) Comparison Across Different Configurations with Nearly Sorted Arrays}
  \Description{Heatmap comparison of Merge Sort and Quick Sort variants.}
  \label{fig:qsvsmsheatmap-nearly-sorted}
\end{figure*}

It is crucial to note that this advantage of optimized Merge Sort is primarily observed with random arrays. For nearly sorted as shown in Figure ~\ref{fig:qsvsmsheatmap-nearly-sorted},  Quick Sort generally remains faster than Merge Sort, even with our optimizations. We observed the same results with sorted arrays, but due to space limitations we were not able to include the figure.


\section{Conclusion}

Our research demonstrates that incorporating fixed sorting networks, like those discovered by AlphaDev~\cite{AlphaDev2023}, into classical sorting algorithms can yield substantial performance improvements.  The 6-to-8 configuration, using networks of sizes 6, 7, and 8, proved particularly effective in optimizing Merge Sort for large arrays, often providing similar benefits to using the full range of available networks (sizes 3 to 8). This highlights the potential of carefully selected sorting networks to significantly enhance the efficiency of traditional sorting methods.

For Merge Sort with sorted arrays, we observed a 2.4x speedup for arrays of size 25,000.  Furthermore, this optimized Merge Sort maintained significant speedups, exceeding 2x, even for larger arrays. This demonstrates the potential of our approach to enhance sorting efficiency across a wide range of input sizes and characteristics.

For Quick Sort, we observe a 1.5x speedup using the 3-to-5 configuration on sorted arrays of size 10,000. The 6-to-8 configuration maintains a consistent 1.5x improvement across sorted arrays from 25,000 to 1 million elements, with the exception of arrays of size 100,000 where 6-to-8 speed up falls a little below 1.5x.

However, our work has some limitations.  One is the CPU architecture dependency of the current sorting networks, which are limited to x86 systems. Future research could focus on discovering or translating these networks for other common CPU architectures to broaden their applicability. Additionally, the limited range of available sorting networks (sizes 3 to 8) restricted our exploration of larger network sizes. Further research could investigate the benefits of larger sorting networks and their integration into classical algorithms.

This work opens new avenues for algorithm optimization, demonstrating the potential of combining classical approaches with AI-driven optimizations. By leveraging the strengths of both, we can achieve significant performance gains and enhance the efficiency of fundamental algorithms like Merge Sort and Quick Sort.

\bibliographystyle{ACM-Reference-Format}
\bibliography{alphadevdnq}

\end{document}